\documentclass[fleqn,12pt]{wlscirep}
\usepackage{amsmath}
\usepackage{graphicx}
\usepackage[utf8]{inputenc}
\usepackage[T1]{fontenc}
\usepackage{lineno}

\usepackage{xcolor}

\newcommand{\aj}{Astron. J.}   
\newcommand{\apj}{Astrophys. J.}   
\newcommand{\apjl}{Astrophys. J. Lett.}   
\newcommand{\apjs}{Astrophys. J. Suppl. Ser.}   
\newcommand{\aap}{Astron. Astrophys.}   
\newcommand{\jcap}{J. Cosmol. Astropart. Phys.}   
\newcommand{\mnras}{Mon. Not. R. Astron. Soc.}   

\newcommand{\nat}{Nature} 
\newcommand{\pasj}{Publ. Astron. Soc. Jpn}   

\newcommand{\arcsec}{^{\prime\prime}}

\usepackage{ulem}  

\title{Reduced incidence of little red dots at z < 3 based on number density and halo mass evolution}

\author[1,2,3]{Chenxuan Zhang}
\author[1,2,3,*]{Huanian Zhang}
\author[1,2,3,*]{Qingwen Wu}
\author[4,5]{Luis C. Ho}
\author[6,7,8]{Jian-Min Wang}

\affil[1]{Department of Astronomy, School of Physics, Huazhong University of Science and Technology, Wuhan, Hubei 430074, China}

\affil[2]{Hubei Fundamental Research Center for Physics, Wuhan, Hubei 430074, China}

\affil[3]{Hubei Key Laboratory of Gravitation and Quantum Physics, Huazhong University of Science and Technology, Wuhan, Hubei 430074, China}

\affil[4]{Kavli Institute for Astronomy and Astrophysics, Peking University, Yiheyuan Road, Beijing, 100871, Beijing, China}

\affil[5]{Department of Astronomy, School of Physics, Peking University, Yiheyuan Road, Beijing, 100871, Beijing, China}

\affil[6]{Key Laboratory for Particle Astrophysics, Institute of High Energy Physics, Chinese Academy of Sciences, Yuquan Road, Beijing, 100049, Beijing, China}

\affil[7]{National Astronomical Observatories of China, Chinese Academy of Sciences, Datun Road, Beijing, 100101, Beijing, China}

\affil[8]{School of Astronomy and Space Science, University of Chinese Academy of Sciences, Yuquan Road, Beijing, 100049, Beijing, China}

\affil[*]{Corresponding authors: huanian@hust.edu.cn; qwwu@hust.edu.cn}

\makeatletter
\renewcommand{\@maketitle}{%
{%
\thispagestyle{empty}%
\vskip-36pt%
{\raggedright\sffamily\bfseries\fontsize{20}{25}\selectfont \@title\par}%
\vskip10pt
{\raggedright\sffamily\fontsize{12}{16}\selectfont  \@author\par}
\vskip25pt%
}%
}%
\makeatother

\begin{document}

\flushbottom
\maketitle

\noindent \textbf{An intriguing puzzle in extragalactic astronomy is the scarcity of Little Red Dots (LRDs) at $z < 3$, compared to their higher abundance at earlier cosmic epochs. To investigate this, we measure the overdensity for 98 specpically confirmed LRDs at $3<z<7$, and find that these LRDs predominantly reside in under-dense regions at $z > 4$  but shift to more typical galaxy environments at $z \sim 3.5$. Concurrently, cross-correlation analyses show that their dark matter halo masses grow rapidly, from $\lesssim 10^{10.1} \, M_{\odot}$ at $z \sim 7.5$ to $\sim 10^{11.3} \, M_{\odot}$ at $z \sim 3.5$, approaching the halo masses of normal galaxies at lower redshifts. Applying an empirical stellar-to-halo mass scaling relation, we find that LRDs still host over-massive black holes relative to their stellar masses at $z > 4$, yet converge toward the local BH--stellar mass relation at lower redshifts. The coherent evolution of LRDs' large-scale environments and halo masses toward those of normal galaxies provides a plausible explanation for their declining abundance at $z < 3$, even though the underlying small-scale physical mechanisms remain elusive.}

The \textit{James Webb Space Telescope} (JWST) has uncovered a previously unrecognized population of compact, red, high-redshift sources, commonly referred to as Little Red Dots (LRDs), distinguished by their unusually steep red rest-frame optical continua, compact size with $\lesssim$ a few $\times 100$ pc, and often unresolved morphologies in NIRCam imaging \cite{labbe2023,harikane2023,furtak2024,greene2024,matthee2024}.
Spectroscopic observations reveal that many LRDs exhibit broad Balmer emission lines, with full widths at half maximum exceeding $\sim1000\ \mathrm{km\ s^{-1}}$, indicative of a central supermassive black hole (SMBH) with inferred mass of order $\sim10^{6}-10^{8}\ M_\odot$ \cite{greene2024,kokorev2024,matthee2024,hviding2025}. 
The V-shaped spectral energy distributions (SEDs) are the most distinctive features of LRDs in the rest-frame UV-to-optical bands, which transit near the Balmer limit \cite{setton2025a,harikane2023,wangbingjie2024}. 
The absorption features on some broad Balmer lines \cite{deugenio2026}, and the absence of X-ray detections \cite{ananna2024,yueminghao2024}, are sometimes attributed to the possible absorption by the dense gas surrounding central SMBHs \cite{inayoshi2025a, inayoshi2025c,jixihan2025}. 
The V-shaped SEDs are frequently explained by the co-existence of AGNs and galaxies with strong dust extinction \cite{chenkejian2025,delvecchio2025,lizhengrong2025}.
However, the lack of direct detections of strong dust emission for high-redshift LRDs appears to be in tension with this interpretation \cite{xiaomengyuan2025,setton2025b}. 
The intrinsically red accretion disk model \cite{zhangchenxuan2026,zwick2026,chen2026} and an envelop dense gas model have been proposed as alternative explanations \cite{inayoshi2025a,inayoshi2025c,jixihan2025}.

The SMBH masses of LRDs, estimated using single-epoch virial methods, are orders of magnitude higher relative to their host galaxy masses than those observed in the local Universe \cite{furtak2024,kocevski2025,tripodi2025,zhangyiyang2025,chenchanghao2025}. 
It should be noted that robust measurements of both SMBH mass and host galaxy stellar mass remain challenging for LRDs, owing to their peculiar observational properties. 
For example, it is still not fully clear that the profile of broad lines in LRDs is caused by virial motion or electron scattering in a dense ionized gas \cite{juodzbalis2026,rusakov2026}. 
Furthermore, no strong extended stellar component has been resolved in these LRDs to date, and the radiation from optical to UV waveband is also controversial (e.g., radiation from either galaxy, AGN disk, or photosphere surrounding a SMBH \cite{mayilun2025,zhangchenxuan2026,wangbingjie2024,jixihan2026}). 
Therefore, precise measurements of both the stellar masses and SMBH masses of LRDs are crucial for understanding the co-evolution of SMBHs and their host galaxies in the early Universe.
Despite being individually faint, LRDs are surprisingly abundant at $z > 4$, with cosmic number densities of $\Phi \sim 10^{-5} - 10^{-4}\ \mathrm{cMpc^{-3}}$ (comoving Mpc) \cite{kokorev2024,akins2025,kocevski2025,matthee2024}, significantly exceeding the space densities of comparably luminous UV-bright quasars at similar epochs. 
The presence of over-massive SMBHs in LRDs implies that either the SMBH growth or the host galaxy assembly proceeds differently from that in other galaxies or quasars.
The environment has long been recognized as an important factor influencing the evolution of galaxies and SMBHs. 
In a recent study, it was found that LRDs may preferentially reside in under-dense regions compared to typical galaxies, based on a sample of photometrically selected LRDs \cite{carranza2025}. 
Thereby, systematic investigations of large-scale environments for a large sample of spectroscopically confirmed LRDs have the potential to shed new light on their distinctive properties and on the disappearance of LRDs at $z<3$ \cite{kocevski2025,mayilun2026,inayoshi2025b}.

In this work, we explore the large-scale environments and halo masses of LRDs based on 98 spectroscopically confirmed LRDs \cite{zhangzijian2026,hviding2025,degraaff2025} at $3<z<7$ identified in JWST/NIRCam and NIRSpec observations across six deep extragalactic fields. 
These LRDs are selected as compact JWST/NIRCam F444W sources exhibiting characteristic V-shaped rest-frame continua, typically characterized by red optical slopes and blue ultraviolet slopes (see Method for more details).
For a subset of these objects, medium- and high-resolution NIRSpec spectra are available, enabling black hole mass estimates from either broad H$\alpha$ or H$\beta$ emission lines using standard virial calibrations \cite{greene2005}. 
To obtain the galaxy sample within the same redshift interval, we draw from the DAWN JWST Archive (DJA) catalogs, which compile robust photometric redshift measurements from multiple public JWST surveys (see Methods for details). 
The galaxy sample is selected to have well-constrained photometric redshifts and to be matched in depth in NIRCam/F200W to the LRDs in each redshift bin, ensuring a fair comparison between the two populations.

\begin{figure*}[!ht]
    \centering
    \includegraphics[width=0.96\textwidth]{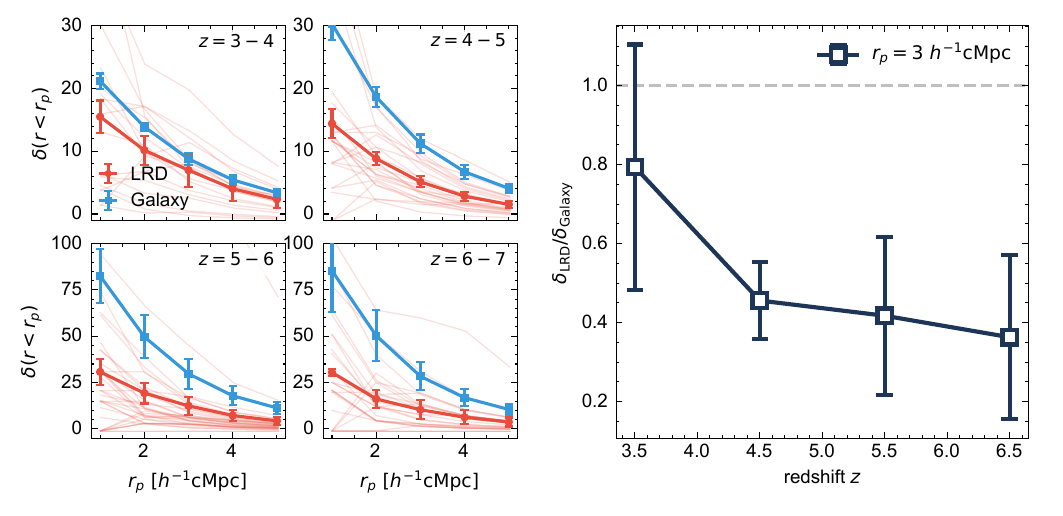}
    \caption{\textbf{Overdensity of LRDs and galaxies across cosmic time.} Left panel: The projected overdensity profiles of LRDs (red) and galaxies (blue) in four redshift bins: $3<z<4$, $4<z<5$, $5<z<6$, and $6<z<7$ with $1\sigma$ uncertainties derived from the jackknife method, where the overdensity for each LRD is also shown with thin red line. Right panel: The ratio of the overdensity of LRDs to galaxies integrated within $r_p < 3\,h^{-1}\,\mathrm{cMpc}$ as a function of redshift.}
    \label{fig:figure_1}
\end{figure*}

To quantify the large-scale environments of LRDs, we measure the galaxy overdensity within cylindrical volumes centered on each LRD, using a line-of-sight velocity window of $\Delta v=\pm2500~\mathrm{km \ s^{-1}}$. 
Due to the limitation of the observing field of view, the maximum projected radii can only reach $\sim$ 5 $h^{-1}$ cMpc. 
The overdensity is defined as $\delta = N_{\rm obs}/N_{\rm exp} - 1$, where $N_{\rm obs}$ is the observed number of galaxies and $N_{\rm exp}$ is the expectation from a random distribution within the same volume. 
The expected counts are estimated using random catalogs that uniformly sample the survey footprints and depths, normalized to the total number of galaxies in each redshift bin.  Similarly, we follow the same procedure to calculate the overdensity for the full galaxy sample. 
We present the overdensity of LRDs and galaxies at $3<z <7$ in Figure \ref{fig:figure_1}, where the error is estimated from the field-level jackknife method (excluding one field each time). Obviously, the overdensity of LRDs is lower than that of galaxies at $z>4$ at all radii and becomes comparable at $3<z<4$. 
This trend extends to the local LRDs at $z\sim 0.1-0.2$ identified in a recent study \cite{linxiaojing2026} (details in Methods). 
The ratio of $\delta_{\rm LRD}$ and $\delta_{\rm Galaxy}$ increases rapidly from $z\sim 6.5$ to $z\sim 3.5$.
This phenomenon demonstrates that LRDs inhabit under-dense regions at $z>4$ and subsequently evolve into environments comparable to those of normal galaxies at $3<z<4$ and even at $z\sim0.2$.

\begin{figure}[!ht]
    \centering
    \includegraphics[width=0.75\linewidth]{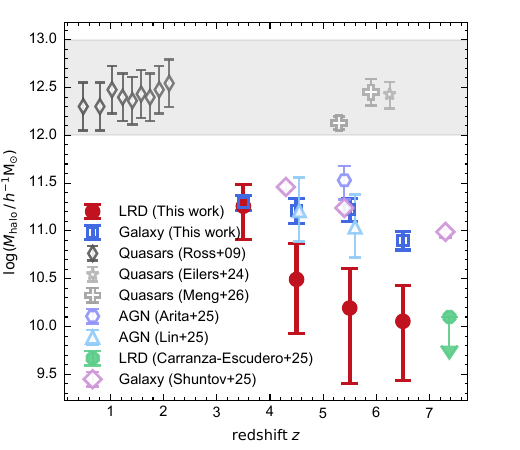}
    \caption{\textbf{Halo masses of LRDs and comparison samples across cosmic time.}
The dark matter halo mass ($M_{\rm halo}$) as a function of redshift for various populations. 
Open blue squares show the galaxies, filled red circles represent LRDs from this work, with downward arrows indicating upper limits. Error bars indicate the 1$\sigma$ uncertainties from the MCMC. Grey symbols denote quasars from clustering analyses \cite{ross2009,eilers2024,menghao2026}. Light blue triangles and violet hexagons show low-luminosity broad-line AGNs from clustering analysis \cite{arita2025,linxiaojinghalo2026}. Green symbols represent LRDs from clustering analysis \cite{carranza2025}. Purple diamonds indicate spectroscopically confirmed massive galaxies \cite{shuntov2025b}. The grey shaded region marks the typical halo mass range ($10^{12}$--$10^{13}\,M_\odot$) for typical Type-1 AGN.}
    \label{fig:figure_2}
\end{figure}

The role of dark matter in galaxy formation and evolution is well established, as it not only shapes the large-scale distribution of galaxies but also, through the large-scale environment, plays a key role in the galaxy assembly process \cite{davis1985,white1978}. 
Clustering analyses, commonly based on the two-point auto-correlation function, have proven to be powerful tools for linking the large-scale properties of galaxies to the small-scale physics, such as their hosting dark matter halos. In this work, we measure the projected cross-correlation function between LRDs and galaxies, as well as the projected autocorrelation function of galaxies in the redshift range of $3 < z < 7$ (see Methods for details).
From these measurements, we infer the dark matter halo masses of galaxies and LRDs, as shown in Figure~\ref{fig:figure_2}. 
We find that the characteristic dark matter halo mass of LRDs grows rapidly from $\lesssim 10^{10.1}\, M_\odot$ at $z \sim 7.5$ to $\sim 10^{11.3}\, M_\odot$ at $z \sim 3.5$, approaching values of typical massive galaxies or AGNs at the same redshift and thus becoming roughly indistinguishable from them.


The evolution of high-redshift galaxies, AGNs, quasars, and their associated dark matter halos has been extensively explored in recent years with JWST.  
Clustering analyses reveal that galaxies with $\langle M_{\rm UV} \rangle \sim -20$, identified as H$\alpha$ or [O {\small} III] emitters, reside in dark matter halos of $\sim 10^{11.5}$ M$_\odot$ at $z\sim 4.3$ and $\sim 10^{11.0}$ M$_\odot$ at $z\sim 7.3$ \cite{shuntov2025b}. 
Low-luminosity AGNs at $3.9 < z < 6$ occupy comparable halos as similarly luminous galaxies, with $\log_{10}(M_{\rm halo, min}/M_\odot) \simeq 11$--$11.5$ \cite{arita2025,linxiaojinghalo2026}. 
In contrast, luminous quasars, whose bolometric luminosities are $\sim 2-3$ orders of magnitude higher than those of such galaxies, low-luminosity AGNs and LRDs, exhibit a minimum host halo mass of $\log_{10}(M_{\rm halo, min}/M_\odot) \simeq 12.4$ at $\langle z \rangle \simeq 6.3$ using five quasar fields \cite{eilers2024}, and a characteristic mass of $\sim 10^{12.2}\, M_{\odot}$ at $5.0 < z \le 5.6$ and $5.6 < z \le 6.2$ from wide-field quasar auto-correlation measurements \cite{menghao2026}.  
Our inferred halo masses for LRDs are lower than those of our galaxy sample and the galaxies in the literature \cite{shuntov2025b} at $4 < z < 7$. 
The halo masses for our general galaxy sample are slightly lower than those reported in the literature \cite{shuntov2025b}, as expected given that our galaxies are slightly fainter in general. 
Furthermore, the derived dark matter halo masses for both galaxies and LRDs are consistent with their measured overdensity, as shown in Figure \ref{fig:figure_1}, as expected from clustering theory.

\begin{figure}[!ht]
    \centering
    \includegraphics[width=1.0\linewidth]{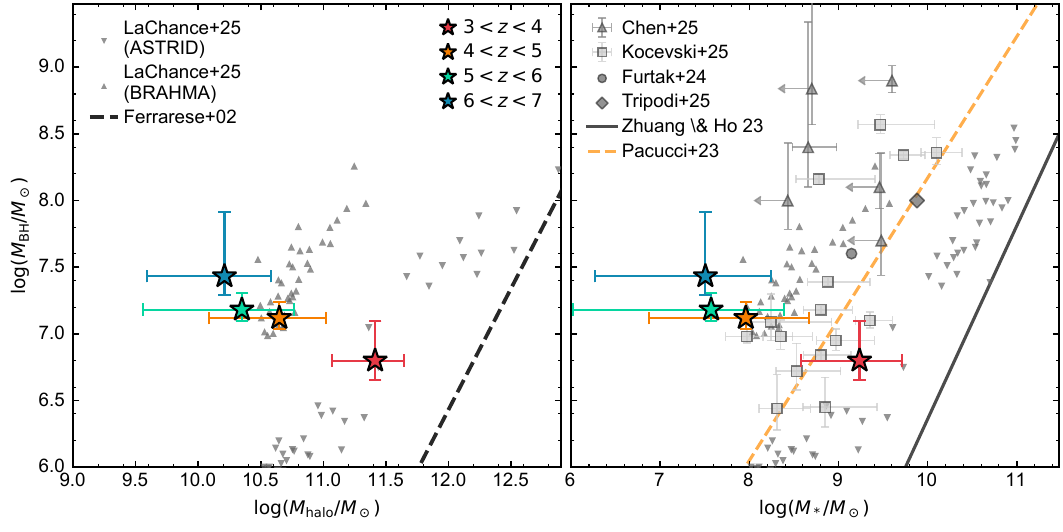} 
    \caption{\textbf{Black hole mass scaling relations for LRDs.}
Solid stars denote the LRDs in our sample, with different colors indicating different redshift bins. 
Leftward arrows indicate upper limits on $M_{\rm halo}$ or $M_*$. 
Upward triangles show LRDs from Chen et al. (2025)\cite{chenchanghao2025}, 
squares represent LRDs from Kocevski et al. (2025)\cite{kocevski2025}, 
downward triangles and leftward triangles show predictions from the {\it ASTRID} and {\it BRAHMA} simulations, respectively \cite{lachance2025}, 
and the circle and diamond mark additional LRDs from Furta et al. (2025) \cite{furtak2024} and Tripodi et al. (2025) \cite{tripodi2025}. 
Horizontal error bars show the $1\sigma$ uncertainties of the halo masses from the MCMC, while vertical error bars denote the $1\sigma$ bootstrap uncertainties of the black hole masses.
In the left panel, the three lines show the local relation \cite{ferrarese2002}. In the right panel, the solid black line shows the local relation  \cite{zhuangmingyang2023}, and the dashed orange line represents the high-$z$ relation  \cite{pacucci2025}.}
    \label{fig:figure_3}
\end{figure}

Investigating the correlations among stellar mass, dark matter halo mass, and supermassive black hole mass is crucial for understanding galaxy evolution and the coevolution of SMBHs and their hosts. In this work, we estimate the average stellar masses of LRDs at a given bin of redshift using the $M_* - M_{\rm halo}$ scaling relation derived from UniverseMachine \cite{behroozi2019}. We present the resulting $M_{\rm halo} - M_{\rm BH}$ and $M_* - M_{\rm BH}$ relation in Figure~\ref{fig:figure_3}. 
We find that LRDs at $z > 4$ deviate significantly from the local relation, whereas LRDs at $z \sim 3.5$ lie much closer to the local correlations of both $M_{\rm halo} - M_{\rm BH}$ \cite{ferrarese2002} and $M_* - M_{\rm BH}$ \cite{zhuangmingyang2023}, despite still exhibiting clear deviations. This result further supports the conclusion that the SMBH masses of LRDs at $z > 4$ are over-massive compared to local galaxies in the $M_{*} - M_{\rm BH}$ relation, and that $M_{*} - M_{\rm BH}$ relation of high-redshift LRDs will evolve to the local relation. Both the $M_{\rm halo} - M_{\rm BH}$ and $M_* - M_{\rm BH}$ relation measured for our LRD sample are well consistent with those in {\it BRAHMA} cosmological simulation at $5 < z < 8$ \cite{lachance2025}, in which the authors produce mock observations of AGNs and their host galaxies to mimic the observational properties of LRDs. In contrast, the {\it ASTRID} cosmological simulation does not reproduce the observational $M_* - M_{\rm BH}$ and/or $M_{\rm halo} - M_{\rm BH}$ relation of LRDs over the same redshift range \cite{lachance2026}. 
It should be noted that the adopted empirical $M_* - M_{\rm halo}$ relation from UniverseMachine may not precisely describe LRDs, whose underlying physical nature remains poorly understood. In addition, black hole mass estimates in LRDs are themselves highly uncertain, as the physical origin of the observed broad emission lines remains debated (e.g., virial motion versus electron scattering \cite{rusakov2026}). If the broad lines are not predominantly virial in origin, the inferred black hole masses could be systematically overestimated by up to $\sim$1 dex. Even under such a reduction, however, the LRDs at $z>4$ in our sample would remain offset above the local $M_{\rm BH}$--$M_{\rm halo}$ scaling relation.


In contrast to luminous AGNs and massive galaxies, which are typically found in over-dense environments \cite{sijacki2009}, LRDs preferentially occupy under-dense regions and are hosted by lower-mass dark matter halos (Figure \ref{fig:figure_2}). This environmental dichotomy suggests that the formation of LRDs may be governed by fundamentally different mechanisms than those of typical galaxies and luminous AGNs, likely shaped by distinct merger and accretion histories \cite{khan2025}. One plausible explanation is that the LRD phase is linked to special halo conditions, such as low halo spins, that become less common or less easily maintained at later times \cite{pacucci2025}. Another is that, as the host galaxies grow in stellar mass and size and experience strong star formation at lower redshift, their spectral energy distributions may naturally evolve out of the color-selection window used to identify LRDs. A third explanation is that the depletion of dense gas at later times could weaken the compact obscured accretion or star-forming phase, thereby significantly altering the spectral energy distributions of LRDs at lower redshifts.

\clearpage

\noindent \textbf{\Large Methods}

\medskip
\noindent {\bf Data Sample}  \label{sec:data}

We construct the LRD sample by combining three recent spectroscopically confirmed JWST LRD catalogs spanning $z\sim2$--9.
One sample is drawn from Zhang et al. (2026)\cite{zhangzijian2026}, which identified 98 LRDs from JWST/NIRCam and NIRSpec observations across six deep fields: A2744, CEERS, GOODS-S, GOODS-N, COSMOS, and UDS \cite{bezanson2024,finkelstein2022,eisenstein2026,dunlop2021}.
These sources were selected as compact NIRCAM/F444W objects exhibiting characteristic V-shaped rest-frame continua, defined primarily by $\beta_{\rm opt}>0$ and $\beta_{\rm UV}<-0.37$, with continuum slopes corrected for emission-line contamination using NIRSpec/PRISM spectroscopy.
The other sample of de Graaff et al. (2025) \cite{degraaff2025}, identified 116 high-purity LRDs from $\sim17{,}000$ high-quality NIRSpec/PRISM spectra using statistically robust V-shaped continuum criteria ($\beta_{\rm opt}>0$, $\beta_{\rm UV}<-0.2$, and $\beta_{\rm UV}-\beta_{\rm opt}>0.5$ at $>95\%$ confidence) together with compact F444W morphologies.
Also, we include the RUBIES spectroscopic sample of Hviding et al. (2025) \cite{hviding2025}, consisting of 36 LRDs identified through joint PRISM+G395M fitting and broad Balmer-line detections, all of which also exhibit compact morphologies and V-shaped continua.
The final merged sample consists of 170 unique LRDs at $2<z<9$ after removing duplicate sources across the three catalogs. 
Restricting the sample to the six deep fields described above, in which the clustering analyses could be conducted, and the redshift range $3<z<7$ yields 98 LRDs used in our analysis. Although the selection criteria differ slightly among the adopted LRD catalogs, we have confirmed that both the environmental measurements and the clustering analyses are mutually consistent, albeit with notably larger uncertainties.

The LRDs' absolute magnitude at 1450 \AA, $M_{1450}$, is derived from the mean spectral flux density measured over $1425-1475$ \AA \  using the DAWN JWST Archive (DJA) public NIRSpec/PRISM dataset (v4) \cite{valentino2023}, as shown in Supplementary Figure~\ref{fig:supp_figure_1}.
Only sources with reliable spectroscopic coverage of the rest-frame UV continuum are included in this measurement.
The majority of the LRD sample is quite faint, and the distribution of $M_{1450}$ v.s. $z$ for our LRDs are consistent with the literature distribution \cite{kocevski2025}. 
The number of LRDs at $z < 3$ and the number of galaxies at $z>7$ are both very limited.  
Therefore, we only focus on LRDs in the redshift intervals $z = 3-4$, $4-5$, $5-6$, and $6-7$, which contain 16, 24, 33, and 25 sources, respectively.  
All LRDs used in the final analysis have spectroscopic redshifts.
For a subset of these sources, NIRSpec medium- and high-resolution grating spectra are available in the DJA Public NIRSpec datasets (v4) \cite{valentino2023}, which are used to estimate the black hole masses by fitting the broad H$\alpha$ or H$\beta$ emission lines and applying the virial mass calibration according to the well-established scaling relation \cite{greene2005}. Although the other possibility that the LRDs are globular clusters in formation is also discussed in a recent study \cite{chisholm2026}. 
We caution that these estimates may be subject to systematic uncertainties associated with the physical origin of the line broadening \cite{rusakov2026}, which could affect the absolute normalization of $M_{\rm BH}$ by $\sim$1 dex.


Similar to a recent study \cite{arita2025}, we construct our galaxy sample at the same redshift range from the DJA catalogs.  
The DJA catalogs are built from publicly available JWST survey data processed by the Cosmic Dawn Center using the \texttt{grizli} \cite{brammer2023a} and \texttt{msaexp} \cite{brammer2023b} pipelines. 
Photometric redshifts are estimated with \texttt{EAZY} \cite{brammer2008} based on combined JWST and HST photometry.
We utilize the DJA v7 catalogs from four survey fields: the Cosmic Evolution Early Release Science Survey (CEERS)\cite{finkelstein2022}, the JWST Advanced Deep Extragalactic Survey (JADES)\cite{eisenstein2026}, the Public Release Imaging for Extragalactic Research (PRIMER) survey  \cite{dunlop2021}, and the Ultradeep NIRSpec and NIRCam Observations Before the Epoch of Reionization (UNCOVER) survey\cite{bezanson2024}.
All UNCOVER galaxies are corrected for gravitational lensing magnification using the v2.0 lensing models \cite{furtak2023,price2025}.

The selection of the galaxy sample is based on two criteria: the uncertainty of photometric redshift estimated using \texttt{EAZY} for each source is less than 0.5, $\delta z_{\rm phot} < 0.5$, and redshift estimates must be based on at least 12 photometric filters ($n_{\rm filter} \geq 12$), to ensure robust redshift estimation. 
As demonstrated in the literature \cite{ye2024}, the photometric redshift estimation is quite reliable and robust.  
Of the 17636 galaxy samples, 727 have spectroscopic redshifts.  
We use spectroscopic redshifts wherever they are available for the galaxy sample. Only when spectroscopic redshifts are unavailable do we rely on photometric redshifts.
The comparison between photometric and spectroscopic redshifts is shown in Supplementary Figure~\ref{fig:supp_figure_2}.
Here we adopt a stricter criterion ($|\Delta z|/(1+z) > 0.1$) for outliers, which account for only 6.60\% of the entire spectroscopic sample, demonstrating the reliability and robustness of photometric redshift estimation. 

Photometry of all sources is measured within a circular aperture with a diameter of $0.5\arcsec$. In each redshift bin, we further restrict the NIRCam/F200W magnitudes of the galaxy sample to be within $\pm 1$ mag of the median F200W magnitude of the LRD sample. This cut is applied to ensure that the effective detection depths of the two populations are comparable, thereby enabling a fair comparison between them. Our galaxy sample lies well within the high-completeness regime of the JWST surveys. As shown in Figure 4 of Merlin et al. (2024) \cite{merlin2024}, the majority of the fields achieve $>90\%$ completeness at $\sim 29$ mag, significantly fainter than the magnitude range considered in our analysis, indicating that incompleteness effects should be minimal for both samples. We additionally confirmed that our main results remain unchanged after excluding the relatively shallow UDS field from the analysis.

A random sample is essential for large-scale clustering analysis. 
We build the random catalogs following Arita et al. (2025)\cite{arita2025}, with a surface number density of $100~{\rm arcmin^{-2}}$ uniformly distributed over the survey regions.  
To accurately trace the survey geometry and depth variations, we retain only random points within the footprint that have $n_{\rm filter} \geq 12$ and NIRCan/F444W observations available.
The redshifts of the random points are assigned to match the redshift distribution of the galaxy sample. The constructed random catalog properly inherits the above properties of the galaxy catalog. 
This random catalog is then used in the estimation of both the overdensity and the two-point correlation function.



We note that there are three local analogs at $z \approx 0.1-0.2$ identified in a recent study \cite{linxiaojing2026}. These objects are identified through template matching to the high-redshift LRD spectra, combined with compact morphology and spectroscopic properties.
To explore the large-scale environments of those three local LRDs, we collect the galaxy sample and the random sample from the large-scale structure (LSS) clustering catalogs, which are constructed from Dark Energy Spectroscopic Instrument (DESI)\cite{desi2016a, desi2016b} Data Release 1 (DR1)\cite{desi2026}. 
The LSS galaxy samples mainly include the Bright Galaxy Sample (BGS) optimized for low-redshift studies \cite{ross2025,adame2025}. 
These catalogs include the observed galaxy positions together with the associated random catalogs, as well as redshifts and survey completeness weights, enabling three-dimensional clustering and environment measurements \cite{ross2025}. 
We also incorporate the Sloan Digital Sky Survey (SDSS) Data Release 12 (DR12) BOSS LOWZ large-scale structure catalog, which consists of luminous red galaxies at $z\lesssim0.4$ and the corresponding random catalog.

Considering that both DESI DR1 and SDSS DR12 around those three local LRDs have low survey coverage, we further supplement the spectroscopic data with the DESI Imaging Legacy Surveys (LS) Data Release 10 (DR10) photometric redshift (photo-$z$) sweeps catalogs, which provide homogeneous photometric redshift estimates for galaxies over the DESI Imaging Legacy  Surveys footprint \cite{dey2019, zhourongpu2023}. 
To ensure consistency with the official LS DR10  photo-$z$ catalog, we apply the recommended selection criteria for the $i$-band extended photo-$z$ sample. Objects are required to have at least one exposure in each of the $g$, $r$, $i$, and $z$ bands ($N_\mathrm{OBS_{G,R,I,Z}} \ge 1$), and only sources with $z$-band magnitude brighter than 21 are considered reliable. 
Additionally, we require a photo-$z$ uncertainty of $\sigma_z < 0.02$ to ensure high-quality redshift estimates. For star-galaxy separation, we exclude point sources (TYPE = `PSF') and apply an optical-WISE color cut ($r - W1 > 0.5$) to preferentially select galaxies, following Zhou et al. (2021) \cite{zhourongpu2021}.
To ensure a consistent comparison, we apply the apparent $g$-band magnitude cut, $20.0<g<21.5$, to both the spectroscopic and photometric samples, to match the apparent magnitudes of the three LRDs.
Also, for the LS DR10 photo-$z$ sweeps catalog, we force the redshift distribution of the random catalog to be matched to that of the real galaxies.

\medskip
\noindent {\bf Environments} \label{sec:environment}

To quantify the large-scale environments of our JWST LRD sample, we calculate the galaxy overdensity, $\delta$, within cylindrical volumes centered on each LRD. 
Specifically, we count galaxies within projected radii of $r_p = 1$, 2, 3, 4, and 5 $h^{-1}$ cMpc and within a line-of-sight velocity range of $\Delta v = \pm 2500$ km s$^{-1}$.  
We have verified that varying the velocity window has almost no effect on the results.
The overdensity is defined as $\delta(r_p) = \frac{N_{\rm obs}(r_p)}{N_{\rm exp}(r_p)} - 1$, where $N_{\rm obs}(r_p)$ is the number of galaxies within the given cylindrical volume, and $N_{\rm exp}(r_p)$ is the expected number of galaxies for the random distribution within the same volume. 
To calculate $N_{\rm exp}$, we count the number of random points within the same cylindrical volume and scale it by the normalization factor $N_{\rm total\ galaxy}/N_{\rm total\ random}$.  
We follow the same procedure to compute the overdensity distribution of the full galaxy sample in the same redshift range.

To estimate the uncertainty in the overdensity parameter $\delta$, we employ a jackknife resampling technique. 
We recompute the mean overdensity $\langle \delta \rangle$ by sequentially excluding one JWST field at a time, yielding $N$ jackknife realizations, where $N$ is the number of fields. 
The jackknife variance is then computed as
\begin{equation}
\sigma^2 = \frac{N-1}{N} \sum_{i=1}^{N} \left( \langle \delta \rangle_i - \langle \delta \rangle \right)^2,
\end{equation}
where $\langle \delta \rangle_i$ is the measurement excluding the $i$-th field. 
We report the square root of this variance as the uncertainty.


To investigate the environments of the three local LRDs, whose SMBH masses are all around $10^{6.5}$ M$_\odot$, we apply the method described above to the DESI and SDSS large-scale structure catalogs. 
Given the limited spectroscopic sky coverage in their vicinity, we measure their overdensity within a volume of (30 $h^{-1}$ Mpc$^3$) to mitigate this constraint, as shown in Supplementary Figure~\ref{fig:supp_figure_3}.
We further complement the spectroscopic data with the LS DR10 photo-$z$ data due to the low sky coverage around these three local LRDs. 
For comparison, the galaxy samples shown in Supplementary Figure~\ref{fig:supp_figure_3} are constructed by randomly selecting 500 galaxies from the DESI DR1 footprint for spectroscopic analysis and an additional 500 galaxies from the LS DR10 catalog for photometric redshift analysis. 
Given the uniform coverage of these surveys and the random selection of galaxy control samples, the overdensity measurements for these control galaxies are not significantly affected by survey incompleteness or edge effects.  
The resulting large-scale environments of these three local LRDs are comparable to those of typical galaxies within the measurement uncertainties.


\medskip
\noindent {\bf Clustering Analysis} \label{sec:clustering}

We evaluate the projected cross-correlation function  between the LRDs and the galaxies of our sample, 
$\omega_{\rm p,CCF}(r_{\rm p})$, and the projected autocorrelation function \cite{peebles1980} of the galaxies, 
$\omega_{\rm p,ACF}(r_{\rm p})$. 
The projected correlation functions are obtained by integrating the 
three-dimensional correlation functions, $\xi_{\rm CCF}(r_{\rm p}, \pi)$ and $\xi_{\rm ACF}(r_{\rm p}, \pi)$, 
where $r_{\rm p}$ and $\pi$ represent the transverse and line-of-sight separations, respectively.

There are various estimators to determine the projected correlation function, and we adopt the Landy \& Szalay estimator \cite{landy1993}, defined as follows:
\begin{equation}
\omega_{\rm p,CCF/ACF}(r_{\rm p}) = \int_{-\pi_{\rm cut}}^{\pi_{\rm cut}} \xi_{\rm CCF/ACF}(r_{\rm p}, \pi)\, d\pi,
\end{equation}
where $\pi_{\rm cut}$ is the maximum line-of-sight separation, fixed to $100\ h^{-1}\mathrm{Mpc}$. 
The estimators for the cross- and auto-correlation functions are given by
\begin{equation}
\xi_{\rm CCF}(r_{\rm p}, \pi)
= \frac{D_{\rm L}D_{\rm G} - D_{\rm L}R_{\rm G} - D_{\rm G}R_{\rm L} + R_{\rm L}R_{\rm G}}
       {R_{\rm L}R_{\rm G}},
\end{equation}

\begin{equation}
\xi_{\rm ACF}(r_{\rm p}, \pi)
= \frac{D_{\rm G}D_{\rm G} - 2D_{\rm G}R_{\rm G} + R_{\rm G}R_{\rm G}}
       {R_{\rm G}R_{\rm G}}.
\end{equation}

Here $D$ and $R$ denote the data and random catalogs, respectively, and the subscripts indicate the corresponding samples.  
The random catalogs are described above. 
All pair counts are normalized by the total number of possible pairs in each sample.
The uncertainties of the projected correlation function are estimated using a leave-one-out jackknife resampling method.
Specifically, for the sample of LRDs, we recompute each time, excluding one LRD from the sample, and derive the uncertainties from the variance among these jackknife realizations.

The projected correlation function is derived from the integration of the real-space correlation function $\xi(r)$\cite{davis1983} as follows:
\begin{equation}
\omega_{\rm p}(r_{\rm p}) = 2 \int_{r_{\rm p}}^{\infty} \frac{r\,\xi(r)}{\sqrt{r^2 - r_{\rm p}^2}}\, dr.
\end{equation}

Assuming a power-law function for the real-space correlation function, $\xi(r) = (r/r_0)^{-\gamma}$, the projected correlation function can be written as
\begin{equation}
\omega_{\rm p}(r_{\rm p})/r_{\rm p} = \mathrm{B}\left(\frac{\gamma - 1}{2}, \frac{1}{2}\right) \left(\frac{r_{\rm p}}{r_0}\right)^{-\gamma},
\end{equation}
where $\mathrm{B}$ denotes the beta function.

Supplementary Figure~\ref{fig:supp_figure_4} shows the projected cross-correlation function between galaxies and LRDs, and Supplementary Figure~\ref{fig:supp_figure_5} displays the projected autocorrelation of galaxies for the different redshift intervals.  The solid lines indicate the best-fit power-law models with $\gamma=1.8$ fixed, and the shaded regions indicate the 1$\sigma$ uncertainty obtained from the Markov Chain Monte Carlo (MCMC) method.

\medskip
\noindent {\bf Halo mass of galaxies and LRDs} \label{halo_mass}

Previous clustering studies have extensively inferred the characteristic dark matter halo masses of quasars over a wide range of redshifts 
\cite{shenyue2007,ross2009,eftekharzadeh2015,timlin2018,hewanqiu2018,arita2023,eilers2024,menghao2026}. 
Here, we estimate the typical dark matter halo (DMH) masses of the LRDs from the clustering measurements obtained above.
The large-scale linear bias parameter quantifies how strongly objects cluster relative to the underlying dark matter distribution. 
We define the bias at a reference scale of $r_{\rm ref}=8\,h^{-1}\,\mathrm{Mpc}$ as
\begin{equation} \label{eq:eq_b}
b = \sqrt{\frac{\xi_{\rm obs}(r_{\rm ref},z)}{\xi_{\rm DM}(r_{\rm ref},z)}},
\end{equation}
where $\xi_{\rm obs}(r,z)$ is the observed correlation function and $\xi_{\rm DM}(r,z)$ is the dark matter correlation function at redshift $z$.

As discussed above, we have adopted the power-law function for the observed correlation function. 
The dark matter correlation function $\xi_{\rm DM}(r,z)$ is computed using the \texttt{halomod}\ package \cite{murray2013, murray2021}. 
We adopt the well-established halo bias model\cite{tinker2010}, a linear matter power spectrum generated with \texttt{CAMB} package\cite{lewis2000}, and the well-known linear growth model\cite{carroll2001}.  
Following the standard relation between auto- and cross-correlation functions\cite{mountrichas2009}, and assuming power-law forms for all correlation functions, we could obtain the linear bias for LRDs:
\begin{equation} \label{eq:ccf_relation}
\xi_{\rm LRD,LRD}(r) = \frac{\xi_{\rm LRD,gal}^2(r)}{\xi_{\rm gal,gal}(r)},
\end{equation}
where $\xi_{\rm LRD,gal}(r)$ is the LRD-galaxy cross-correlation function and $\xi_{\rm gal,gal}(r)$ is the galaxy autocorrelation function, from which we could derive the effective correlation length $r_{0,\rm LRD}$ and the bias parameter $b_{\rm LRD}$ for LRDs. The halo mass corresponding to a given bias parameter is obtained by inverting the halo bias $-$ halo mass relation. 
We use the \texttt{colossus} implementation of the well-developed halo bias model \cite{tinker2010} with a halo mass definition of $M_{200\mathrm{m}}$. The clustering results are presented in Supplementary Table~\ref{tab:supp_halo_mass_clustering}.

Recent clustering studies have provided constraints on the typical halo masses of LRDs, but the inferred values show significant variations among different samples and methodologies. For example, Zhuang et al. (2025) inferred relatively massive host halos ($M_{\rm halo}\gtrsim10^{12} M_\odot$) from the angular clustering of 14 bright LRDs, while Pan et al. (2026) derived lower halo masses ($M_{\rm halo}\sim$ a few $\times10^{11} M_\odot$) from a sample of 34 spectroscopically confirmed LRDs \cite{zhuang2026, pan2026}. The differences among these measurements may arise from a combination of factors, including sample selection, luminosity distribution, redshift distribution, and clustering methodology. In particular, current samples remain relatively small and span different regions of the LRD parameter space, making direct comparisons challenging. Larger spectroscopic samples with homogeneous selection will therefore be essential to establish robust constraints on the typical environments and halo masses of LRDs.

\medskip
\noindent {\bf The $M_\ast$ of LRDs} \label{sec:sedfit}

To infer the stellar masses corresponding to our halo mass measurements, we adopt the redshift-dependent stellar-to-halo mass relation (SMHM) from UniverseMachine \cite{behroozi2013, behroozi2019}. More discussions on SMHM can be found in a recent study \cite{romeo2020}. 
This formalism parametrizes the median stellar mass as a function of halo mass and redshift via the following equation:
\begin{equation}
\log M_\ast(M_{\rm h},z) = \log(\epsilon M_1) + f\!\left(\log\frac{M_{\rm h}}{M_1}\right) - f(0),
\end{equation}
where the function $f(x)$ is defined as:
\begin{equation}
f(x) = -\log\left(10^{\alpha x} + 1\right) + \delta \frac{\left(\log\left(1 + e^x\right)\right)^\gamma}{1 + e^{-x}},
\quad x = \log\frac{M_{\mathrm{h}}}{M_1},
\end{equation}
with $\alpha$ controlling the low-mass slope ($M_* \propto M_{\mathrm{h}}^\alpha$ at low halo masses), and $\delta$ and $\gamma$ jointly regulating the curvature and amplitude at the high-mass end. 
These shape parameters $(\alpha, \delta, \gamma)$, along with the characteristic halo mass $M_1$ and normalization $\epsilon$, evolve with redshift as given by Equations (3)--(5) in the paper \cite{behroozi2013} and $M_1$ and $\epsilon$ represent the characteristic halo mass and normalization, respectively. 

We note that although UniverseMachine was calibrated using pre-JWST observations, the SMHM relation remains robust across the mass range we explored. Recent JWST studies \cite{shuntov2022,shuntov2025c} suggest a higher SMHM ratio at $z \gtrsim 4$ at the high-mass end, implying enhanced star-formation efficiency in the early Universe. However, for the halo mass range relevant to our sample ($M_{\rm halo} < 10^{11.5}\,M_\odot$ for $z<5$, and $M_{\rm halo} \lesssim 10^{11.0}\,M_\odot$ for $z>5$), the difference remains modest, with stellar masses differing by only $0.1-0.2$ dex even at $z \sim 7$.

\subsection*{Data Availability}

The reduced JWST data underlying this study are publicly available through the DAWN JWST Archive (DJA; \url{https://dawn-cph.github.io/dja/}). 
All raw JWST imaging and spectroscopic data used in this work are publicly available from the Mikulski Archive for Space Telescopes (MAST) under their respective survey program IDs.
Specifically, the JWST data were obtained from the following public programs and survey fields: 
the JWST Advanced Deep Extragalactic Survey (JADES; Program IDs 1181 and 1210), 
The Public Release Imaging for Extragalactic Research survey (PRIMER; Program ID 4233), 
The Cosmic Evolution Early Release Science Survey (CEERS; Director’s Discretionary Early Release Science Program ID ERS-1345), 
and the Ultradeep NIRSpec and NIRCam Observations Before the Epoch of Reionization survey (UNCOVER; Program ID GO-2561).
The DESI DR1 LSS catalogs and Imaging Legacy Survey Data Release 10 catalogs are publicly available at \url{https://data.desi.lbl.gov}.
The SDSS DR12 data products are publicly available at \url{https://data.sdss.org/sas/dr12/boss/lss/}.

\subsection*{Code Availability}

All software packages used in this analysis are publicly available. 
Specifically, \texttt{halomod} \cite{murray2013,murray2021}, 
\texttt{CAMB} \cite{lewis2000}, and \texttt{colossus} \cite{tinker2010} can be obtained from their respective public repositories.

\subsection*{Acknowlegements}
We acknowledge Linhua Jiang and Zijian Zhang at the Kavli Institute for Astronomy and Astrophysics, Peking University, for providing the LRD sample.

\subsection*{Funding Statement}
This study was supported by the National Natural Science Foundation of China (NSFC, grants No. 12533005, 12233007 and 12303007), the China Manned Space Project (CMS-CSST-2025-A06, CMS-CSST-2025-A07), the National Key Research and Development Program of China (No. 2023YFC2206702) and the start-up funding of Huazhong University of Science and Technology.

\subsection*{Author Contributions}
HZ designed the project. CZ led the data analysis, and HZ provided guidance. CZ, HZ, and QW led the manuscript writing. QW and HZ led the interpretation of the results. All authors (CZ, HZ, QW, LH, JW) contributed to the discussion and manuscript revision.

\subsection*{Competing interests} 
The authors declare no competing interests.


\clearpage
\section*{Main Figures} 

\clearpage

\clearpage

\clearpage
\setcounter{figure}{0}
\setcounter{table}{0}
\renewcommand{\figurename}{Supplementary Figure}
\renewcommand{\tablename}{Supplementary Table}
\renewcommand{\thefigure}{\arabic{figure}}
\renewcommand{\thetable}{\arabic{table}}
\ifdefined\theHfigure
  \renewcommand{\theHfigure}{supp.\arabic{figure}}
\fi
\ifdefined\theHtable
  \renewcommand{\theHtable}{supp.\arabic{table}}
\fi
\section*{Supplementary Information}

\begin{figure}[!ht]
    \centering
    \includegraphics[width=0.7\linewidth]{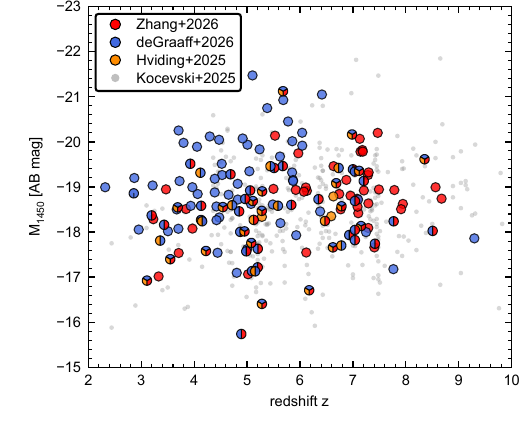}
    \caption{\textbf{Redshift and absolute magnitude distributions of LRDs.} The distributions of redshift and absolute magnitude at 1450 \AA \ ($M_{1450}$) for LRDs in the redshift range of $2.0 < z < 10$ are shown. Larger colored circles indicate the LRDs in our sample compiled from multiple studies \cite{zhangzijian2026,hviding2025,degraaff2025}, while small grey dots represent LRDs compiled from 
    Kocevski et al. (2025) \cite{kocevski2025}.}
    \label{fig:supp_figure_1}
\end{figure}
\clearpage


\begin{figure}[!ht]
    \centering
    \includegraphics[width=0.6\linewidth]{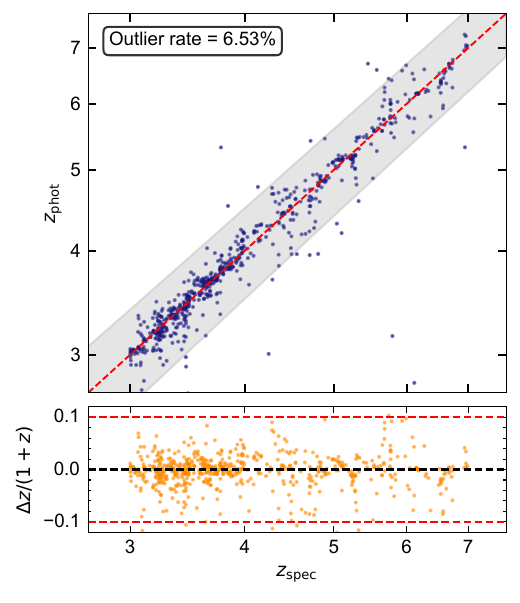}
    \caption{\textbf{Photometric versus spectroscopic redshift for galaxies.} Top panel: Comparison between photometric redshift ($z_{\rm phot}$) and spectroscopic redshift ($z_{\rm spec}$) for the galaxy sample across redshift range of $z = 3-7$. The dashed red line displays the 1:1 relation, and the grey shaded region denotes $|\Delta z|/(1+z) < 0.1$. Bottom panel: Normalized redshift residuals $\Delta z/(1+z) = (z_{\rm phot} - z_{\rm spec})/(1+z_{\rm spec})$ as a function of spectroscopic redshift. Horizontal dashed lines mark $\Delta z/(1+z) = 0$ (black) and $\pm 0.1$ (red). The outliers are defined as sources with $|\Delta z|/(1+z) > 0.1$.}
    \label{fig:supp_figure_2}
\end{figure}
\clearpage

\begin{figure}[!ht]
    \centering
    \includegraphics[width=0.8\linewidth]{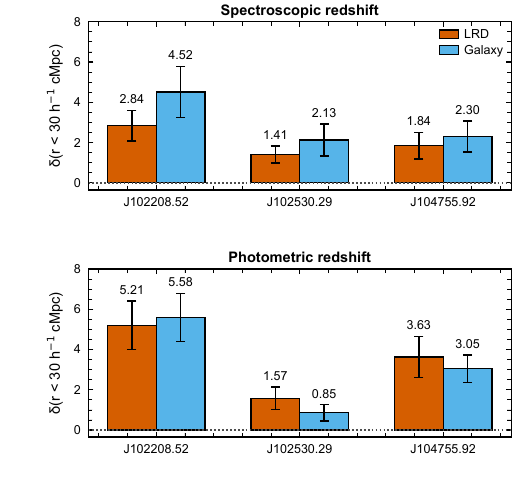}
    \caption{\textbf{Overdensity of the three local LRDs and galaxies within a volume of (30 $h^{-1}$ cMpc)$^3$.} Top panel: overdensity measurements within a volume of (30 $h^{-1}$ cMpc)$^3$ around the three local LRDs (coral red bars) and the galaxy sample (blue bars), using spectroscopic data from the combined SDSS and DESI surveys. Bottom panel: overdensity measurements using the photometric redshift catalog. Error bars represent $1\sigma$ uncertainties derived from jackknife resampling for LRDs and galaxies, and the numbers are the values of the overdensity measurements. In both panels, the x-axis labels correspond to the three local LRD identifiers.}
    \label{fig:supp_figure_3}
\end{figure}
\clearpage

\begin{figure*}[!ht]
    \centering
    \includegraphics[width=0.96\linewidth]{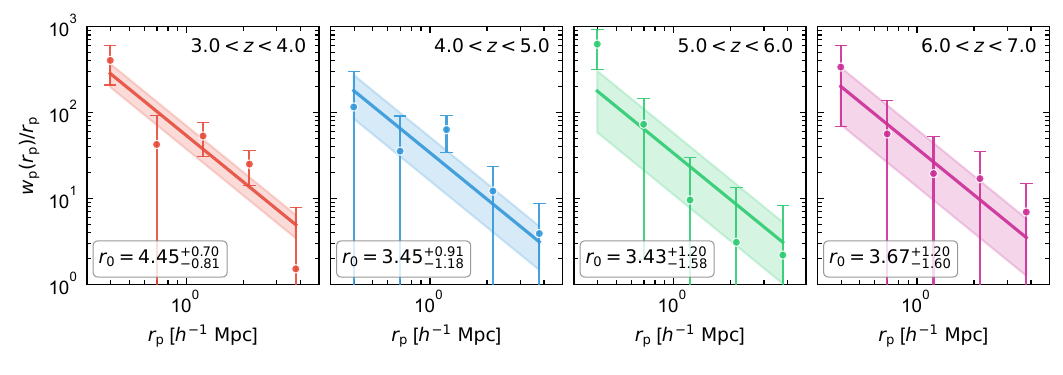}
    \caption{\textbf{Projected cross-correlation functions between LRDs and galaxies.} 
    The projected cross-correlation function $\omega_{\rm p}(r_{\rm p})$ as a function of projected separation $r_{\rm p}$ for four redshift bins: $3.0 < z < 4.0$ (left), $4.0 < z < 5.0$ (second from left), $5.0 < z < 6.0$ (second from right), and $6.0 < z < 7.0$ (right). 
    Colored circles with error bars represent the measured projected cross-correlation function between LRDs and surrounding galaxies.
    $1 \sigma$ error is obtained from the jackknife method. 
    Solid lines show the best-fit power-law models with $\gamma = 1.8$, and shaded regions indicate the $1\sigma$ uncertainty in the correlation length $r_0$. 
    The best-fit correlation lengths are displayed in each panel.}
    \label{fig:supp_figure_4}
\end{figure*}
\clearpage

\begin{figure*}[!ht]
    \centering
    \includegraphics[width=0.96\linewidth]{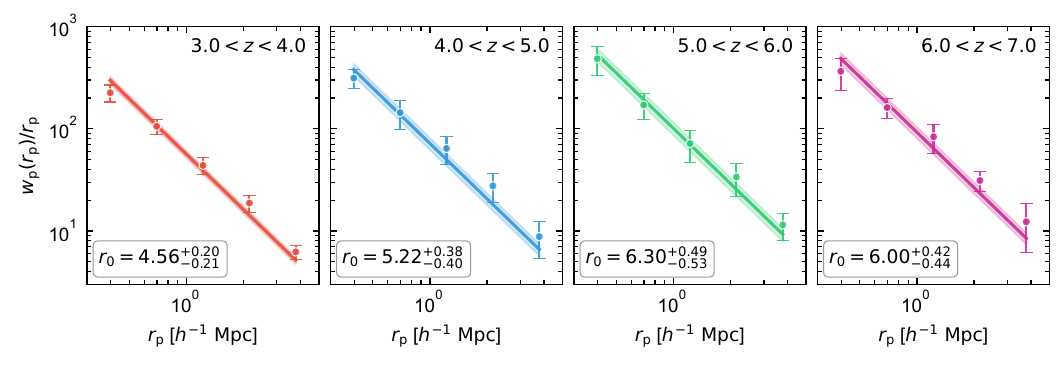}
    \caption{\textbf{Projected auto-correlation function of galaxies.} 
    The projected auto-correlation function $\omega_{\rm p}(r_{\rm p})$ as a function of projected separation $r_{\rm p}$ for four redshift bins: $3.0 < z < 4.0$ (left), $4.0 < z < 5.0$ (second from left), $5.0 < z < 6.0$ (second from right), and $6.0 < z < 7.0$ (right). 
    Colored circles with error bars represent the measured projected auto-correlation function of galaxies.
    $1 \sigma$ error is obtained from the jackknife method. 
    Solid lines show the best-fit power-law models with $\gamma = 1.8$, and shaded regions indicate the $1\sigma$ uncertainty in the correlation length $r_0$. 
    The best-fit correlation lengths are displayed in each panel.
    }
    \label{fig:supp_figure_5}
\end{figure*}

\clearpage
\subsection*{Supplementary Table} 

\begin{table*}[!ht]
\centering
\small
\setlength{\tabcolsep}{4pt}
\renewcommand{\arraystretch}{1.4}
\caption{\textbf{Best-fit parameters.} Best-fit correlation lengths, galaxy bias factors, and inferred dark matter halo masses measured from galaxy auto-correlation functions (ACF) and LRD--galaxy cross-correlation functions (CCF) in different redshift bins. $N_{\rm LRD}$ and $N_{\rm galaxy}$ denote the number of LRDs and photometric galaxies used in each bin. The reported uncertainties correspond to $1\sigma$ confidence levels derived from the MCMC sampling.}

\label{tab:supp_halo_mass_clustering}
\resizebox{\linewidth}{!}{%
\begin{tabular}{ccccccccc}
\hline
Redshifts & $N_{\rm LRD}$ & $N_{\rm galaxy}$ & $r_{0,\rm CCF}$ & $r_{0,\rm ACF}$ & $b_{\rm galaxy}$ &
$b_{\rm LRD}$ & $\log(M_{\rm halo, galaxy})$ & $\log(M_{\rm halo, LRD})$\\
 & & & ($h^{-1}\,\mathrm{Mpc}$) & ($h^{-1}\,\mathrm{Mpc}$) & & & $(h^{-1}\rm M_\odot)$ & $(h^{-1}\rm M_\odot)$\\
\hline
$3-4$ & 16 & 6431 & $4.45^{+0.70}_{-0.81}$ & $4.56^{+0.20}_{-0.21}$ & $3.01^{+0.13}_{-0.12}$ & $2.94^{+0.40}_{-0.47}$ & $11.30^{+0.07}_{-0.07}$ & $11.26^{+0.23}_{-0.34}$ \\
$4-5$ & 24 & 5408 & $3.45^{+0.91}_{-1.18}$ & $5.22^{+0.38}_{-0.40}$ & $4.17^{+0.28}_{-0.28}$ & $2.85^{+0.83}_{-0.64}$ & $11.22^{+0.11}_{-0.12}$ & $10.49^{+0.43}_{-0.46}$ \\
$5-6$ & 33 & 2550 & $3.43^{+1.29}_{-1.58}$ & $6.30^{+0.49}_{-0.53}$ & $5.84^{+0.41}_{-0.44}$ & $3.35^{+1.10}_{-1.23}$ & $11.23^{+0.11}_{-0.13}$ & $10.19^{+0.41}_{-0.70}$ \\
$6-7$ & 25 & 1293 & $3.67^{+1.20}_{-1.60}$ & $6.00^{+0.42}_{-0.44}$ & $6.41^{+0.39}_{-0.41}$ & $4.11^{+1.26}_{-1.33}$ & $10.90^{+0.10}_{-0.11}$ & $10.05^{+0.43}_{-0.58}$ \\
\hline
\end{tabular}
}
\end{table*}

\newpage




\clearpage


\end{document}